\documentclass[12pt]{article}
\usepackage{graphicx}
\usepackage{lscape}
\usepackage{rotating}
\usepackage{natbib}
\bibpunct{(}{)}{;}{a}{}{,}
\usepackage{amssymb,amsmath}
\usepackage{epsfig}
\usepackage{color} 
\usepackage{txfonts}

\usepackage{threeparttable} 
\usepackage{lscape}

\begin{document}

\title{Ultra-High-Energy Cosmic Ray Contribution from the Spin-Down Power of Black Holes}
 \maketitle
 \begin{center}
Ioana Du\c{t}an\footnote{Member of the International Max Planck Research School (IMPRS) for Astronomy and Astrophysics at the Universities of Bonn and Cologne. Member of the Pierre Auger Collaboration.}\\

\begin{small}
Max Planck Institute for Radio Astronomy, Auf dem H\"{u}gel 69, 53121 Bonn, Germany\\
Research Center for Atomic Physics and Astrophysics, Str. Atomi\c{s}tilor, nr. 405, Bucharest-M\v{a}gurele, Romania\\ 
email: \texttt{idutan@brahms.fizica.unibuc.ro}\end{small}

\section*{Abstract}
\end{center}
We investigate the production of ultra-high-energy cosmic ray (UHECR) in jets from low-luminosity active galactic nuclei (LLAGN). We propose a model for the UHECR contribution from the spin-down power of black holes (BHs) in LLAGN, which present a jet power $P_{\mathrm{j}} \leqslant 10^{46}$ erg s$^{-1}$. This is in contrast to the opinion that only high-luminosity AGN can accelerate particles to energies $ \geqslant 50$ EeV. We rewrite the equations which describe the synchrotron self-absorbed emission of a non-thermal particle distribution to obtain the observed radio flux density from flat-spectrum core sources and its relationship to the jet power. In general, the jet power provides the UHECR luminosity and so, its relationship to the observed radio flux density. We found that the UHECR luminosity is dependent on the observed radio flux density, the distance to the AGN, and the BH mass, where the particle acceleration regions can be sustained by the magnetic energy extraction from spinning BHs and where the strength of the magnetic field at the sites of particle acceleration scales with the maximum value of the BH magnetic field, which is $\sim 10^4$ gauss for a BH of $10^9 M_{\odot}$. We apply the model to M87 and Cen  A, two possible sources of UHECRs, whose jet parameters can be inferred from observational data. Next, we use a complete sample of 29 steep spectrum radio sources with a total flux density greater than 0.5 Jy at 5 GHz to make predictions for the maximum particle energy, luminosity, and flux of the UHECRs from nearby AGN. Using our proposed model, it is possible to show that LLAGN can be sites of the origin of UHECRs. In additional, the scenario in which the contribution to the UHECR flux from many weak radio galaxies would dominate over that from a few strong radio galaxies, or vice-versa, should be substantiated with further statistics.

\section{Introduction}

Cosmic rays (CRs) are a direct sample of matter from outside the solar system, and their study can, for instance, provide important information on the chemical evolution of the universe or improve constraints on Galactic and extragalactic magnetic fields. They can be measured indirectly through the study of extensive air showers that are induced as the CRs hit the top of the atmosphere (known as CR events). The extensive air showers are currently observed using air fluorescence [e.g., High Resolution Fly's Eye (HiRes) experiment\footnote{http://www.cosmic-ray.org}] or large array, ground-based detectors [e.g.,  Akeno Giant Air Shower Array (AGASA){\footnote{http://www-akeno.icrr.u-tokyo.ac.jp/AGASA}], or both [e.g., Pierre Auger Observatory (Auger)\footnote{http://www.auger.org}]. In the future, space-based detectors might be another option. UHECR particles are mostly protons or fully ionized nuclei with energy above 50 EeV (1 EeV = 10$^{18}$ eV). At such high energies, the flux of UHECRs is very low and only a few dozen particles per square kilometer per century are expected. This is one of the main reasons for the difficulty posed in understanding the origin and nature of the UHECRs. Therefore, very large detector arrays are required. The Pierre Auger Observatory, by far the biggest cosmic ray detection instrument, uses air fluorescence and water detection in a hybrid instrument with an aperture of 7000 km$^2$ sr.

Joint efforts have been made during the past decade by worldwide, cosmic ray experiments to help us understand from where the UHECRs come and what is their nature. It is believed that the UHECRs originate in extragalactic sources, as the gyroradius of a proton with an energy of 100 EeV is of the order of the dimension of our galaxy, whereas most of the CR particles with energy below 50 EeV originate within our galaxy \cite[e.g.,][]{berezinsky06,stanev10b,stanev10}. If the UHECR particles are protons, they are subject to energy loss by creating pions through their occasional collisions with the cosmic microwave background (CMB) photons. This process produces a suppression of the cosmic ray energy spectrum beyond 50 EeV, which is known as the Greisen-Zatsepin-Guzmin (GZK) cutoff \citep{greisen,zatsepin-kuzmin}. Therefore, the UHECRs would not be able to survive the propagation from their acceleration sites to us unless their sources are located within $\sim 100$ Mpc. The presence of the GZK cutoff at the expected energy in the data released by the HiRes collaboration was taken as strong evidence that the UHECR flux is dominated by protons \citep{data2}. 

A suppression of the CR flux has also been observed in the data released by the Pierre Auger collaboration \citep{spectrum,data1}. With respect to primary composition, this collaboration has exploited the
observation of the longitudinal shower development with fluorescence detectors to measure the depth of the maximum of the shower evolution, $X_{\rm max}$, which is sensitive to the primary mass. A gradual increase of the average mass of cosmic rays with energy up to 59 EeV is deduced when comparing the absolute values of $X_{\rm max}$ and RMS($X_{\rm max}$) to air shower simulations \citep{data1}.

The present data collected by the Auger Collaboration, which consists of 69 events of energy $E\geqslant 55$ EeV, shows an anisotropy in the arrival direction of the UHECRs \citep{auger07,auger08,auger10b}. Moreover, the arrival direction of the UHECRs is statistically correlated with the distribution of nearby extragalactic objects [AGN and gamma-ray bursts (GRBs)], where the region around the position of the radiogalaxy Cen A has the largest excess of arrival directions relative to the isotropic expectations.

At highest energies, proton propagation is affected only by the CMB, whereas heavy nuclei may be deflected by Galactic magnetic fields \cite[e.g.,][]{medina98}.

UHECRs are most probably accelerated at astrophysical shocks, for instance, through a first-order Fermi mechanism \cite[e.g.,][]{ga99}, in very powerful systems that can be associated with jets and hot spots in AGN and GRBs. The magnetic field plays an important role for the particle acceleration mechanism. The field should be strong enough to confine the particles in the acceleration region, but at the same time, weak enough to avoid too much loss by radiative cooling. Such shocks can be associated with Poynting flux models for the origin of jets from force-free magnetosphere above thin accretion disks, which were first proposed by \citet{lovelace} and \citet{bland76}. In the model by Lovelace, the accreting protons are accelerated in the potential drop across the accretion disk by electric forces, which then form two high-current, aligned, and opposite proton beams. The output electrical power in the beams is proportional to the maximum energy of the protons squared, $L \sim E^2_{\mathrm{max}}$. The maximum energy to which the accretion disk can accelerate the proton beams is set by the Eddington luminosity. If one takes the Poynting flux as a lower limit to the energy flux along a relativistic jet, UHECR production in LLAGN cannot be explained. \citet{biermann08} rewrote Lovelace's equation as
\begin{equation}
L_{\mathrm{P}} =\frac{c}{4\pi} f_{\mathrm{flare}} \left( \frac{E_{\mathrm{max}}}{eZ\gamma_{\mathrm{sh}}}\right)^2,
\label{Pflux}
\end{equation} 
where  $Z$ is the mass number of the nuclei, $\gamma_{\mathrm{sh}}$ is the Lorentz factor of the shock, and $f_{\mathrm{flare}} (< 1)$ is the intermittency. As the authors state, probably all three elements are required if one considers UHECR production by sources like M87 and Cen A, whose energy flow along the jet are $< 10^{45}$ erg s$^{-1}$ and $< 10^{43}$ erg s$^{-1}$, respectively \citep{why-antonucci}. 

As an alternative, \citet{Farrar09} showed that very intense, short-duration AGN flares that result from the tidal disruption of a star or from a disk instability can accelerate UHECRs. On the other hand, magnetic reconnection in relativistic jets represents another option for UHECR acceleration \cite[e.g.,][]{giannios10}.

\citet{boldt-ghosh} suggested that particles with energies $\geqslant 10^{20}$ eV may be accelerated near the event horizons of spinning BHs associated with presently inactive quasar remnants. The required electromagnetic force is generated by the BH induced rotation of externally supplied magnetic field lines threading the horizon, where the magnetic field is supported by external current and the horizon is an imperfect conductor with resistance $\sim 100\,\Omega$. Therefore, the BH behaves as a battery, driving currents around a circuit, with an electromagnetic force of up to $10^{21}$ eV for a BH with a mass of $10^{9} M_{\odot}$ \cite[e.g.,][]{lovelace}. In this case, the production of observed flux of the highest energy cosmic rays would constitute a negligible drain on the BH dynamo. That is, replenishing the particle ejected at high energies ($> 10^{20}$ eV) would require a minimal mass input; a luminosity of $10^{42}$ erg s$^{-1}$ in such particles (if protons) corresponds to a rest mass loss $< 10^{-5} M_{\odot}$ in a Hubble time. Particle acceleration to UHEs from the spin-down power of BHs was also discussed in \citet{blandford00}, where the acceleration regions can be sustained by magnetic energy extraction from spinning BHs.

In this paper, we propose a model for UHECR contribution from the spin-down power of BH in LLAGN. The particles in the jet manage to tap the spin-down power of the BH and then are accelerated at relativistic shocks with energies up to the UHE domain. The electrons lose their energy through synchrotron emission, whereas the protons are capable of surviving the radiative cooling and perhaps of propagating through the intergalactic and Galactic medium towards us. Since both particles undergo the same acceleration process, there must be a correlation between the electron synchrotron emission and the UHECR proton energy. We seek this correlation to make predictions for maximum energy, luminosity, and flux of the UHECRs from nearby LLAGN. This is in contrast to the opinion that only high-luminosity AGN can accelerate particles to UHE domain \cite[e.g.,][]{zaw09}.

In Section \ref{model}, we provide a description of the model. In Section \ref{jetpower}, we derive the relation between the jet power and the observed radio flux density for a flat-spectrum core source. Based on this relation, we derive in Section \ref{lumUHECR} the luminosity and flux of the UHECRs. In Section \ref{emax}, we calculate the particle maximum energy taking into account the spatial limit and synchrotron emission losses. In Section \ref{m87cenA}, we present the application of the model to M87 and Cen A. Both sources are LLAGN with a mass accretion rate relative to the Eddington accretion rate less than $\sim 10^{-2}$ (therefore, they can be powered by the BH spin down) and present strong radio-emitting jets. In Section \ref{sources}, we provide the predictions for nearby galaxies as possible sources of UHECRs. In Section \ref{sec:summary}, we present a summary of the key points and discuss the implication of this model for further studies of the UHECRs.

\section{Model description}
\label{model} 

\subsection{Model conditions}
\begin{itemize}
\item We assume that the UHECRs are accelerated by shocks in AGN jets, which are launched from the inner accretion disk which is located inside the BH ergosphere \citep{eu04}. The inner disk extends from the stationary limit $r_{\mathrm{sl}}$ inward to the innermost stable orbit $r_{\mathrm{ms}}$. When the mass accretion rate relative to the Eddington accretion rate is $ \sim 10^{-2}$, the jets can be powered by the spin down of the BH, which is transferred to the disk inside the ergosphere by closed magnetic field lines that connect the BH to the accretion disk. The jet propagates along a cylinder of length $z_{\mathrm{0}}$ (see Fig.~\ref{fig:jetGeom}) using the BH spin-down power and then extends into a conical shape with a constant opening angle $2\,\theta$, as a consequence of the free adiabatic expansion of the jet plasma. (The tip of the cone is located at some $z < z_{\mathrm{0}}$.) A similar geometry of the jet was considered by \citet{markoff01}.
\item The calculations are performed for the case when the UHECRs would have been protons. For heavy nuclei of a given atomic number ($Z$), the particle energy will scale up with $Z$.
\item In the observer frame, the magnetic field along the jet varies as $B \sim \gamma_{\mathrm{j}}^{-1}z^{-1}$ and the electron number density in the jet scales as $ \sim \gamma_{\mathrm{j}} z^{-2}$, where $\gamma_{\mathrm{j}}$ is the bulk Lorentz factor of the jet. (See discussions below.)
\item We set the slope of the particle density distribution to $p = 2$, which corresponds to a flat-spectrum core sources with a spectral index $\alpha = 0.5$, and the strength of the BH magnetic field to its maximum value ($B_{\rm H}^{\rm max}$). The latter condition provides, in turn, the minimum values of the particle maximum energy, luminosity, and flux of the UHECRs. In Section \ref{m87cenA}, we discuss some implication of a different choice for the value of $p$.
\end{itemize}

\begin{figure}\centering
\epsfig{file=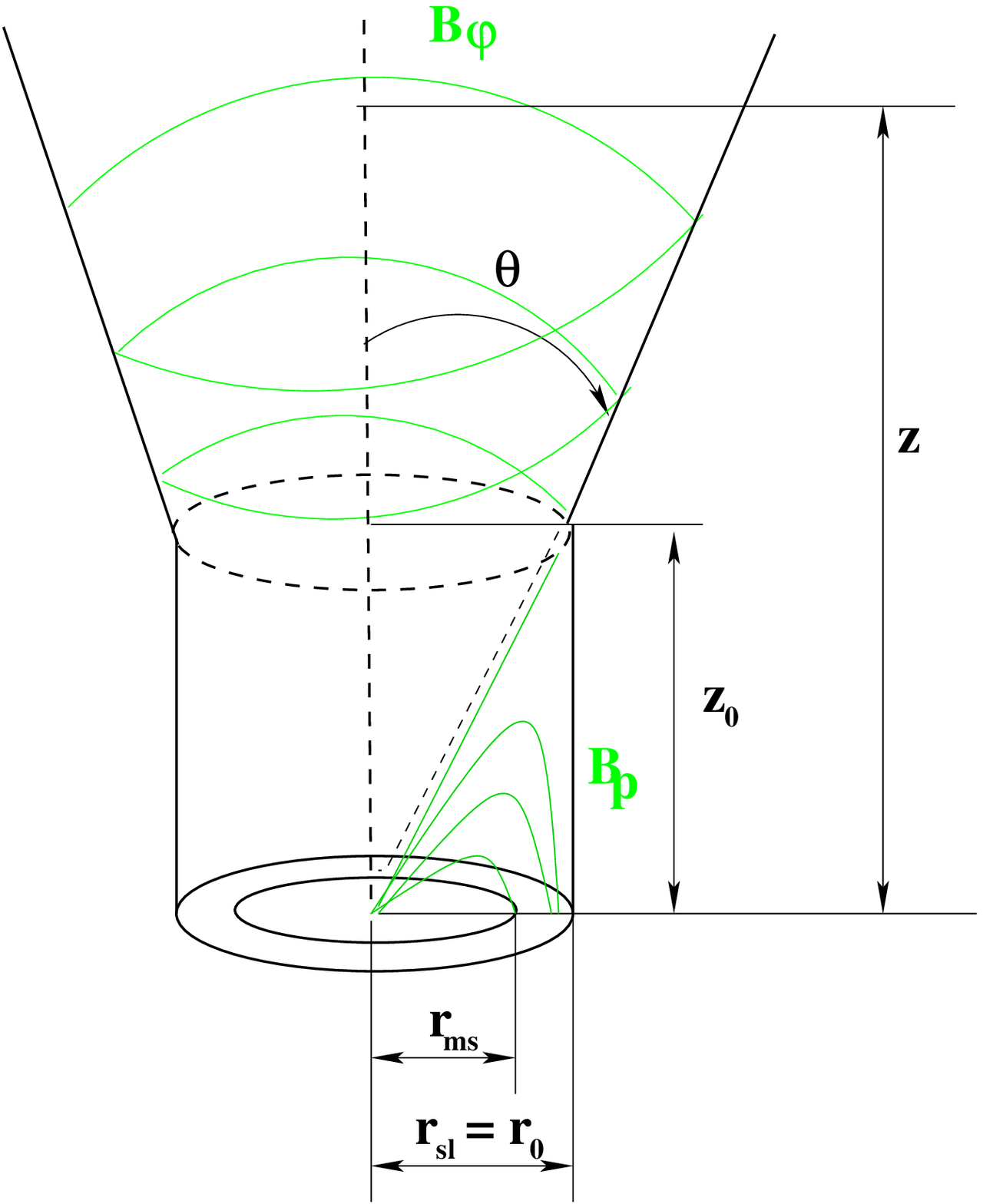,height=5.5cm}
\caption{Schematic representation of the jet geometry. The jet is launched from the inner disk, extending from the stationary limit inward to the innermost stable orbit, and propagates along a cylinder up to a distance of $z_{\mathrm{0}}$. It then expands freely into a conical geometry with a constant opening angle $2\,\theta$. The magnetic field lines threading the disk near the BH (dashed lines) are wound up, far from the BH, into a toroidal magnetic field $B^{\phi}$ that collimates the jet.}
\label{fig:jetGeom}
\end{figure}

\subsection{\label{magn}Magnetic field scaling along a steady jet}

To describe the jet physics, we use the following reference frames: (i) the frame comoving with the jet and (ii) the (rest) frame of the observer, in which the relativistic jet moves with the bulk Lorentz factor.

In a frame comoving with the jet, the poloidal component of the magnetic field is considered to vary as $B_{\mathrm{p}} \sim z^{-2}$. This variation follows from the conservation of magnetic flux along the axis $z$. To keep the field divergence-free, the toroidal component must vary as $B_{\mathrm{\phi}} \sim z^{-1}$. This topology of $B_{\mathrm{\phi}} \sim z^{-1}$ was first derived by \citet{parker} for the magnetohydrodynamics solution of a spherical-symmetric flow (so that, a jet can be considered a conical cut along the flow surfaces). [See also \citet{bk79}.] The observational support to this variation of $B_{\mathrm{\phi}}$ is specified later on in this section. 
At a distance, say, $z_0$, the poloidal and toroidal components of the comoving magnetic field become approximately equal $B_{\mathrm{p0}} \simeq B_{\mathrm{\phi0}}$. We consider $z_0$ of a few gravitational radii,\footnote{The gravitational radius is defined as $r_{\mathrm{g}} \equiv GM/c^2=r^{\dagger}_{\mathrm{g}}(M/10^9 M_{\odot})=1.48\times 10^{14}(M/10^9 M_{\odot})$ cm, where $G$ is the Newtonian gravitational constant, $M$ is the BH mass, and $c$ is the speed of light.} based on the fact that the VLBI observation, for instance, of the jet in M87 at 7 mm gives evidence on the jet collimation (by the toroidal magnetic field) on scales of 60-200 $r_{\mathrm{g}}$ \citep{biretta02} and the global 3.5 mm VLBI observations have resolved sizes for the compact radio sources of $\sim 10\,   r_{\mathrm{g}}$ \cite[e.g.,][]{lee}. A large-scale and predominantly toroidal magnetic field can exert an inward force (hoop stress), confining and collimating the jet \cite[e.g.,][]{bisnovatyi-ruzmaikin,bp82}. The magnetic hoop stress is balanced either by the gas pressure of the jet or by centrifugal force if the jet is spinning. From $z_0 $ upward, the poloidal component of the magnetic field becomes weaker, so that the field lines are soon wound up in the azimuthal direction by the jet rotation. Thus, above $z_0$, the magnetic field along the jet is nearly azimuthal $B \sim B_{\phi}$ (for a steady jet) and varies inversely proportional to the distance along the jet:
\begin{equation}
B = B_0 \left( \frac{z}{z_0}\right)^{-1} , 
\label{magnfield2}
\end{equation}
where $B_0 \equiv B_{\mathrm{\phi0}} \simeq B_{\mathrm{p0}}$ is the strength of the magnetic field at the height $z=z_0$ above the equatorial plane of the BH. This $z$-dependence of the magnetic field appears to be contradicted by radio-polarization observations \citep{bridle&perley}. These observations strongly suggest that the magnetic field is predominantly parallel to the jet axis initially and only later becomes perpendicular to the jet axis, with some parallel magnetic field left over. \citet{pbb09} argued that the basic pattern of the magnetic field is indeed $B_{\phi} \sim z^{-1}$ and that the observational evidence for a parallel magnetic field is due to highly oblique shocks. Their argument is based on the observations of the jet structure which indicate the occurrence of the moving shocks between 20 and 200 $r_{\mathrm{g}}$, while the first stationary, strong shock can be produced in the approximate range of $(3 - 6) \times10^{3}$ $r_{\mathrm{g}}$ \citep{markoff01,markoff05,marscher}.

The strength of the magnetic field in the comoving frame $B_0$ can be related to the poloidal magnetic field in the BH frame $B_{\mathrm{H}}$ \cite[e.g.,][]{drenkhahn} as
\begin{equation}
B_0=\frac{1}{\gamma_ {\mathrm{j}}}B_{\mathrm{H}}=\frac{B_{\mathrm{H}}^{\mathrm{max}}}{\gamma_ {\mathrm{j}}}\left(\frac{B_{\mathrm{H}}}{B_{\mathrm{H}}^{\mathrm{max}}} \right) ,
\label{field}
\end{equation} 
where the maximum value of the BH magnetic field is given by
\begin{equation}
B_{\mathrm{H}}^{\mathrm{max}} \simeq  0.56 \times 10^4 \left(\frac{M}{10^9M_{\odot}}\right) ^{-1/2}\ \mathrm{gauss},
\label{maxB}
\end{equation} 
which is obtained in a similar manner as the calculation performed by \citet{znajek78}, with the difference that we set the BH potential drop to the specific energy of the particles at the innermost stable orbit, whereas \citet{znajek78} makes use of the fact that the Eddington luminosity sets an upper bound on the radiation pressure (as the disk is radiatively efficient). The maximum value of the BH magnetic field corresponds to the time when the accretion rate was as high as the Eddington accretion rate. In this case, the BH spin parameter\footnote{The BH spin parameter is defined as $a_* \equiv J/J{\max}\, (= a/r_{\mathrm{g}})$, where $a=J/Mc$ is the angular momentum of the BH about the spinning axis per unit mass and per speed of light and $J_{\max} = GM^2/c$ is the maximal angular momentum of the BH. Furthermore, the BH spin parameter obeys the condition: $-1\leq a_*\leq +1$. } is limited to $a_*=0.9982$ \citep{t74}. Although this limit might be even closer to the maximal value of the spin parameter $\sim 1$, this will introduce just a small variation of the maximum value of the BH magnetic field.

\subsection{Electron and proton number densities}
The jet is assumed to be composed mainly of electrons, positrons, and protons. We denote by  $f_{\mathrm{ep}} \equiv n_{\mathrm{e}}/n_{\mathrm{p}}$ the ratio of the electron to proton number densities, where the number densities are measured in a frame comoving with the jet plasma. Unless otherwise noted, $n_{\mathrm{e}}$ should be assumed to include the positron number density as well. It is straightforward to generalize to a mixed chemical composition, including many heavy nuclei. Furthermore, both electrons and protons can have thermal and non-thermal populations before being accelerated at the shock. There may also be a substantial number of positrons from pion production and decay processes (also called secondaries).

Now, we look for the expression of the proton and electron number densities injected into the accelerating region. First, we consider the mass flow rate into the jets, which in the comoving frame is given by
\begin{equation}
\dot{M}_{{\mathrm{j,co}}} =\frac{d}{dt}\left( \rho_{\mathrm{j}} V_{\mathrm{j}} \right) =\frac{d}{dt} \left[ n  m  z (S)_{z=0} \right] = n  m  v_{\mathrm{j}}(S)_{z=0} ,
\label{mjet}
\end{equation} 
where $\rho_{\mathrm{j}} $ is the rest-mass density of the jet, $V_{\mathrm{j}} $ is the comoving volume of the jet, $(S)_{z=0}$ is the launching area of the jet, $z$ is the length of the cylinder along which the jet propagates before expanding freely in a conical geometry, and  $v_{\mathrm{j}} = \beta_ {\mathrm{j}} c $ is the bulk velocity of the jet.

The surface area between two equatorial surfaces of a Kerr BH can be calculated as 
\begin{equation}
(dS)_{z=0}=\left( \frac{A}{\Delta}\right) ^{1/2}2\pi dr ,
\end{equation} 
where the Kerr metric functions are:
\begin{equation}
\Delta =r^2-2r_\mathrm{g}r+a^2 \ \mathrm{and}\ A=r^4+r^2a^2+2r_\mathrm{g}ra^2 ,
\label{coeff}
\end{equation}
where $r$ is the coordinate radius. Next, we use normalizations to the gravitational radius, so that $r_*=r/r_{\mathrm{g}}$ is the dimensionless radius. The surface area is then:
\begin{equation}
(S)_{z=0} = 2\pi r_{\mathrm{g}}^2 \int_{r_{\mathrm{ms}_*}}^{r_{\mathrm{sl}_*}}\limits r_* \sqrt{\frac{1 + r_*^{-2}a_*^2 + 2r_*^{-3}a_*^2}{1 - r_*^{-1} + r_*^{-2}a_*^2}}dr_* \equiv  2\pi r_{\mathrm{g}}^2 k_0 ,
\label{aria}
\end{equation} 
where the factor $k_0$ increases from $\sim$ 2 to $\sim$ 80 as the BH spin parameter increases from 0.95 to $\sim$ 1. For the first equality, we use the fact that the inner disk, from where the jet is launched, has its inner and outer radii at the innermost stable orbit\footnote{Once the accretion flow reaches the innermost stable orbit, it drops out of the disk and falls directly into the BH. The expression for the radius of the innermost stable orbit $r_{\mathrm{ms}}$ is given by eq. (2.21) in \citet{bardeen70}.} $r_{\mathrm{ms}}$ and stationary limit $r_{\mathrm{sl}}= 2 r_{\mathrm{g}} \equiv r_0$, respectively. 

The comoving density of the jet can be expressed in terms of the ratio of the electron to proton number densities:
\begin{equation}
n  m=n_{\mathrm{p}} m_{\mathrm{p}} + n_{\mathrm{e}} m_{\mathrm{e}} = n_{\mathrm{p}} m_{\mathrm{p}} \left( 1+ f_{\mathrm{ep}} \frac{m_{\mathrm{e}}}{m_{\mathrm{p}}}\right) \equiv n_{\mathrm{p}} m_{\mathrm{p}}f_0 .
\label{density}
\end{equation} 
For protons dominating over the electrons, $f_{\mathrm{ep}} < 2\times10^{3}$, where electrons and positrons can partially occur as secondaries. This fraction of CR is $\sim 10^{-2}$ from data of CRs at 1 GeV. One can get a ratio of unity assuming that the spectra go down to rest mass, which is implausible [see, e.g., \citet{protheroe96} and references therein].

Substituting Eqs. (\ref{aria}) and (\ref{density}) for (\ref{mjet}), we obtain the mass flow rate into the jet in the observer frame by including $\gamma_{\mathrm{j}}$:
\begin{equation}
\dot{M}_{{\mathrm{j}}} =\gamma_{\mathrm{j}} \beta_ {\mathrm{j}} c   n_{\mathrm{p}} m_{\mathrm{p}} f_0   2 \pi r_{\mathrm{g}}^2 k_0.,
\label{eq:mdot}
\end{equation} 
This expression provides the proton number density, which we use to derive the electron number density:
\begin{equation}
n_{\mathrm{e}}=f_{\mathrm{ep}} \frac{\dot{M}_{{\mathrm{j}}}}{\gamma_{\mathrm{j}} \beta_ {\mathrm{j}} c   m_{\mathrm{p}} f_0   2 \pi r_{\mathrm{g}}^2 k_0 }.
\label{edensity}
\end{equation} 
We shall use this result later for evaluating the self-absorbed synchrotron emission of the jets (Section \ref{secSAS}).

\subsection{Particle energy distribution}

We suppose that a shock is produced at the jet height $z = z_0$ \cite[e.g.,][]{markoff01}. As a result, a power-law energy distribution of the particles is established. For a given frequency, the emission from the synchrotron process for electrons and protons gives a factor $(m_{\mathrm{p}}/m_{\mathrm{e}})^{3}\simeq 10^{10}$  in favor of electrons \cite[e.g.,][]{nt359}. In addition, \citet{bs} showed that the proton synchrotron emission can be competitive if one considers that the proton emission ranges to much higher photon energy. Moreover, it is also not at all obvious that they have the same normalization at the same Lorentz factor of the particle and that the particles have a continuous power-law from rest mass to UHE. The number density of the electrons in the energy interval $E$, $E + dE$  [or $m_{\mathrm{e}}c^2\gamma$, $m_{\mathrm{e}}c^2(\gamma + d\gamma$)] has the power-law form:
\begin{equation}
N(E) dE=C E^{-p} dE
\label{distrib}
\end{equation}
or, in terms of the Lorentz factor,
\begin{equation}
N(\gamma) d\gamma=C' \gamma^{-p} d\gamma ,
\label{distrib2}
\end{equation} 
where $\gamma\in[\gamma_{\mathrm{min}},\gamma_{\mathrm{max}}]$ is the Lorentz factor of the electrons and $p$ is the power-law index \cite[e.g.,][]{ryb+ligh}. The normalization coefficients of the electron number density in Eqs. \ref{distrib} and \ref{distrib2} are related by 
\begin{equation}
C=C' (m_{\mathrm{e}}c^2)^{p-1} .
\label{norm}
\end{equation}
The normalization of the electron number density $C'$ follows the pressure in a reheating flow ($ \sim z^{-2}$), whereas an adiabatic flow would give a steeper dependence, which leads to shocks \cite[e.g.,][]{sanders}. Adiabatic behavior implies $P\sim \rho^{\gamma_{\mathrm{ad}}}$, where $\gamma_{\mathrm{ad}}$ is the adiabatic index. Since a relativistic fluid usually has $\gamma_{\mathrm{ad}} = 4/3$, in an adiabatic flow the temperature runs as $z^{-2/3}$. Furthermore, one can have a conical flow only if the temperature of the flow is approximately constant. The energy for reheating can be taken from the flow through highly oblique shocks. The normalization of the electron number density, in the case of a conical jet, is therefore:
\begin{equation}
C'=C'_0\left( \frac{z}{z_0}\right)^{-2}  (\textrm{cm}^{-3}) .
\label{norm2}
\end{equation} 

Possible values of the power-law distribution index $p$ of the electrons accelerated by the relativistic shock are discussed later in Section \ref{m87cenA}.

\subsection{Self-absorbed synchrotron emission of the jets}
\label{secSAS}

The spectra from compact radio sources can be explained by self-absorbed synchrotron emission of the jets produced by electrons with a power-law energy distribution. In this section, we rewrite the quantities which describe the self-absorbed synchrotron emission, derived in \citet{ryb+ligh}, and express them under the considerations of the model presented here. We first introduce the absorption coefficient, optical depth, synchrotron emissivity, and source function in order to calculate the flux density of the synchrotron emission from radio sources with a flat-spectrum core. (The quantities which describe the self-absorbed synchrotron emission are in cgs units.)

In a frame comoving with the jet plasma, the absorption coefficient of the synchrotron radiation can be calculated as
\begin{equation}
\begin{split}
\alpha_{\nu} = \: & \frac{\sqrt{3} e^3 B\sin\alpha_0}{8\pi m_{\mathrm{e}}}\left( \frac{3 e B\sin\alpha_0}{2\pi m_{\mathrm{e}}^3  c^5}\right)^{\frac{p}{2}} C \\
& \Gamma\left( \frac{3p+2}{12}\right) \Gamma\left( \frac{3p+22}{12}\right) \nu^{-\frac{p+4}{2}}\;(\textrm{cm}^{-1}) ,
\end{split}
\end{equation} 
where $e$ is the electron electric charge, $m_{\mathrm{e}}$ is the electron mass, $p$ is the power-law index of the particles distribution, $C$ is the normalization factor for the power-law electron energy distribution (Eq. \ref{distrib}), $B$ is the magnetic field in the frame comoving with the jet, $\Gamma(x)$ is the Gamma function of argument $x$, and $\nu$ is the frequency of the synchrotron radiation [eq. 6.26 in \citet{ryb+ligh}]. The average over the pitch angle $\alpha_0$, for a local randomly oriented magnetic field with a probability distribution $\frac{1}{2}\sin\alpha_0   d\alpha_0$, is given by the integral in \citet{longair}. Including the values of the physical constants and using the expressions for the normalization of the electron distribution function, $C$ and $C'$ (Eqs. \ref{norm} and \ref{norm2}), as well as for comoving magnetic field along the jet (Eq.~\ref{magnfield2}), the absorption coefficient then becomes:
\begin{equation}
\alpha_{\nu} = K_1 C'_0   \left( \frac{z}{z_0}\right)^{-\frac{p+6}{2}}    B_0^{\frac{p+2}{2}}   \nu^{-\frac{p+4}{2}} ,
\label{absorptionFin}
\end{equation} 
where
\begin{equation}
\begin{split}
K_1 = \: & 8.4\times10^{-3} (1.25\times10^{19})^{\frac{p}{2}}  \left( 8.2\times 10^{-7}\right)^{p-1} \frac{\sqrt{\pi}}{2} \\
&\Gamma\left( \frac{3p+2}{12}\right)  \Gamma\left(  \frac{3p+22}{12}\right)   \Gamma\left( \frac{p+6}{4}\right)  \Gamma^{-1}\left(  \frac{p+8}{4}\right).
\end{split}
\end{equation} 

To calculate the observed distance along the jet where the jet becomes self-absorbed, we first determine the optical depth $\tau_{\nu}$ of the jet material. The averaged path of a photon through the jet has the length $r(z)$, which is a reasonable approximation for a jet observed at large inclination angle \cite[e.g.,][]{kaiser}. We introduce a factor $l_{0}$ in the expression of the path length to account for a small inclination angle. Thus, we can write the optical depth as
\begin{equation}
\tau_{\nu}=\alpha_{\nu} r(z)l_{0} .
\end{equation} 
For conical jets, the intrinsic half-opening angle is given by $\tan\theta=r/z \cong r_0/z_0$. With the absorption coefficient specified through Eq. (\ref{absorptionFin}), the optical depth can be written as
\begin{equation}
\tau_{\nu} = K_1  C'_0  r_0  l_0  \left( \frac{z}{z_0}\right) ^{-\frac{p+4}{2}} B_0^{\frac{p+2}{2}} \nu^{-\frac{p+4}{2}} ,
\label{tau}
\end{equation} 

One can define the distance along the jet where the jet becomes self-absorbed $z_{ssa}$ as the distance $z$ for which $\tau_{\nu}=1$. Using Eq. (\ref{tau}), one obtains:
\begin{equation}
z_{ssa} = \left(  K_1  C'_0  l_0\right) ^{\frac{2}{p+4}}\left(\tan\theta \right)^{-1}  r_0^{\frac{p+6}{p+4}} B_0^{\frac{p+2}{p+4} \nu^{-1}} .
\label{zet}
\end{equation} 

The total power radiated per unit volume per unit frequency by a non-thermal particle distribution equals:
\begin{equation}
\begin{split}
P_{\omega}=  \: & \frac{\sqrt{3} e^3}{2\pi  m_ {\mathrm{e}}c^2}\frac{C' B \sin\alpha_0}{p+1} \left( \frac{m_{\mathrm{e}} c \omega}{3 e B \sin\alpha_0}\right)^{-\frac{p-1}{2}}\\
&\Gamma\left(\frac{p}{4}+\frac{19}{12} \right)\Gamma\left(\frac{p}{4}-\frac{1}{12} \right) ,
\end{split}
\end{equation} 
where $\omega=2\pi\nu$ [Eq. 6.36 in \citet{ryb+ligh}]. Using Eqs. (\ref{magnfield2}) and  (\ref{norm2}), as well as the method to calculate the averaged pitch angle employed in \citet{longair}, the expression of the total power becomes:
\begin{equation}
P_{\nu} = 2\pi P_{\omega}= K_2   C'_0   \left( \frac{z}{z_0} \right)^{-\frac{p+5}{2}} B_0^{\frac{p+1}{2}} \nu^{-\frac{p-1}{2}} ,
\label{Ptot}
\end{equation}  
where
\begin{equation}
\begin{split}
K_2 = \: & 3.7\times10^{-23}  \left( 1.2\times10^{-7}\right) ^{-\frac{p-1}{2}}(p+1)^{-1}\frac{\sqrt{\pi}}{2} \\ 
& \Gamma\left(\frac{p}{4}+\frac{19}{12} \right)\Gamma\left(\frac{p}{4}-\frac{1}{12} \right)\Gamma\left(\frac{p+5}{4}\right)\Gamma^{-1}\left(\frac{p+7}{4}\right) .
\end{split}
\end{equation}
Next, we derive $(z/z_0)$ from Eq. \ref{tau} when $\tau_{\nu}=1$. With this, the expression for the total power takes the form:
\begin{equation}
P'_{\nu} = K_2 \left(K_1  r_0  l_0\right)^{-\frac{p+5}{p+4}} (C'_0)^{-\frac{1}{p+4}} B_0^{-\frac{p+3}{p+4}} \nu^3,
\end{equation} 
and the emission coefficient is simply $j_{\nu} = P_{\nu}/4\pi$. 

The emission coefficient is defined as the product between the absorption coefficient $\alpha_{\nu}$ and the source function $S_{\nu}$. At low frequencies, the emitting region is opaque to synchrotron radiation and the observed intensity of radiation $I_{\nu}$ is proportional to the source function, while at high frequencies, the region is transparent and the observed intensity is proportional to the emission coefficient. The two dependences should be matched at the transition from opaque to transparent regimes. This transition corresponds to an optical depth $\tau_{\nu} = 1$.  The source function in the self-absorbed limit is then:
\begin{equation}
S_{\nu} = \frac{1}{4\pi}\frac{P_{\nu}}{\alpha_{\nu}} = \frac{1}{4\pi}\frac{K_2}{K_1}\left(\frac{z}{z_0} \right)^{\frac{1}{2}}B_0^{-\frac{1}{2}} \nu^{\frac{5}{2}} \;(\textrm{erg s}^{-1}\textrm{cm}^{-2}\textrm{Hz}^{-1}) ,
\end{equation} 
where the last equality was obtained using Eqs. (\ref{absorptionFin}) and (\ref{Ptot}). Note that in the source function, the emitting frequency does not depend on the power-law index of the electron energy distribution. For $\tau_{\nu}=1$, the source function becomes:
\begin{equation}
S'_{\nu}=K_3  \left( C'_0  r_0  l_0\right)^{\frac{-1}{p+4}}  B_0^{-\frac{1}{p+4}}\nu^2 \;(\textrm{erg s}^{-1}\textrm{cm}^{-2}\textrm{Hz}^{-1}) ,
\label{snu}
\end{equation} 
where $K_3 = K_1^{-\frac{p+3}{p+4}}K_2$.

To obtain the emission spectrum, one needs to solve the equation for the radiative transfer through a homogeneous medium. Because the angular sizes of the jets are small, instead of the specific intensity of the radiation, one usually measures the flux density $F_{\nu}$ (energy per unit time, per unit frequency interval, that passes through a surface of unit area). Thus,
\begin{equation}
dF_{\nu} = I_{\nu}d\Omega = S_{\nu}\left[ 1-\exp(-\tau_{\nu})\right] d\Omega.
\label{df}
\end{equation} 
Because the frequency shift of the approaching photons, specified by the Doppler factor\footnote{The Doppler factor of the jet is $\mathcal{D}_{\mathrm{j}}=\gamma_{\mathrm{j}}^{-1}(1-\beta_{\mathrm{j}}  \cos\varphi)^{-1}$, where $\varphi$ is the inclination angle of the jet axis with respect to the line of sight (which is Lorentz transformed through $\sin{\varphi}_{\mathrm{obs}}= \mathcal{D}_{\mathrm{j}}\sin{\varphi}$). The angles are rotated by the Lorentz transformation, so that a jet seen at angle $\gamma_{\mathrm{j}}^{-1}$ is rotated basically to a transverse view for large $\gamma_{\mathrm{j}}$.}, is $\nu_{\mathrm{obs}} =  \mathcal{D}_{\mathrm{j}}\nu$, the transformation of the specific intensity to the observer frame is:
\begin{equation}
I_{\nu,\mathrm{obs}} = \mathcal{D}_{\mathrm{j}}^{3}I_{\nu} ,
\label{ints}
\end{equation} 
where the relativistic invariant quantity $I_{\nu}/\nu^{3}$ was used. The solid angle corresponding to the source is:
\begin{equation}
d\Omega =\frac{2\pi r dz}{D_{\mathrm{s}}^2} ,
\label{omega2}
\end{equation} 
where $D_{\mathrm{s}}$ is the distance from the observer to the jet source and $r= z \tan\theta$. If we insert Eq. (\ref{omega2}) into Eq. (\ref{df}) and integrate it from $z_0$ to $z$, we obtain the flux density of the synchrotron emission in the case of 
$\tau_{\nu}=1$ as
\begin{equation}
F'_{\nu} = S'_{\nu}[1-\exp(-1)]  \pi (\tan\theta) D_{\mathrm{s}}^{-2} z^2\left[1-\left(\frac{z_0}{z} \right)^{2} \right] ,
\end{equation} 
where the second term in the last squared bracket can be neglected with respect to the first term for $z \gg z_0$ (where the jet emission becomes self-absorbed). Using Eqs. (\ref{zet}) and (\ref{snu}), the flux density is then: 
\begin{equation}
F' = K_4  (C'_0  l_0)^{\frac{5}{p+4}}   r_0^{\frac{2p+13}{p+4}} B_0^{\frac{2p+3}{p+4}} D_{\mathrm{s}}^{-2}  (\tan\theta)^{-1},
\label{flux}
\end{equation} 
where $K_4 = 0.16   K_1^{-\frac{p-1}{p+4}}K_2$. The radio flux density in Eq. (\ref{flux}) does not depend on the emitted frequency of the radiation since we already adopted the case of flat-spectrum core sources when $\tau_{\nu}=1$. 

For a power-law synchrotron spectrum of the form $F_{\mathrm{obs}}\sim \nu_{\mathrm{obs}}^{-\alpha}$, the observed flux density is related to the intrinsic flux density as
\begin{equation}
F_{\mathrm{obs}}=\mathcal{D}_{\mathrm{j}}^{3+\alpha} F' ,
\label{blob}
\end{equation} 
where Eq. (\ref{ints}) was used. Equation (\ref{blob}) is valid for a single blob emission \cite[e.g.,][]{bk79}. For a continuous jet consisting of uniformly-spaced blobs, one needs to consider the emission per unit length in the observer frame. Therefore,
\begin{equation}
F_{\mathrm{obs}}=\mathcal{D}_{\mathrm{j}}^{2+\alpha} F' ,
\label{cblobs}
\end{equation} 
due to the fact that the number of the blobs observed per unit length is $\sim 1/\mathcal{D}_{\mathrm{j}}$.

\section{Relation between the jet power and the observed radio flux density for a flat-spectrum core source}
\label{jetpower}

In the previous section, we established the expression for the radio flux density from flat-spectrum core sources (Eq. \ref{flux}). This quantity reflects the radiative property of the jet, as the radiated energy is replaced by dissipation of the jet power \citep[e.g.,][]{bk79}. In this section, we seek the relation between the jet power and the observed radio flux density. First, we consider the jet power in the observer frame defined as
\begin{equation}
P_{\mathrm{j}} =  \gamma_ {\mathrm{j}}{\dot{M}}_{\mathrm{j}}c^2,
\label{jet}
\end{equation} 
which follows, e.g., from \citet{fb95} [see also \citet{vila10}], and for which we need to evaluate ${\dot{M}}_{\mathrm{j}}$ using Eq. \ref{edensity}. An upper limit for the electron density is specified by $n_{\mathrm{e}} \leqslant C'_0 $. So, we can substitute Eq. (\ref{edensity}) for the expression of the observed radio flux density (either Eq.~\ref{blob} or Eq.~\ref{cblobs}, depending on the structure of the jet) and find the mass flow rate into the jet ${\dot{M}}_{\mathrm{j}}$. The strength of the magnetic field $B_0$ follows from Eqs. (\ref{field}) and (\ref{maxB}). This procedure yields the power of the jet: 
\begin{equation}
\begin{split}
P_{\mathrm{j}} =  & K_5 f   \beta_ {\mathrm{j}}\mathcal{D}_{\mathrm{j}}^{-h} \left( \frac{\gamma_ {\mathrm{j}}}{5}\right) ^{\frac{2p+13}{5}} \left( \frac{\tan\theta}{0.05}\right)^{\frac{p+4}{5}} \left( \frac{r_0}{2r_{\mathrm{g}}}\right) ^{-\frac{2p+13}{5}}  \\
&\left(\frac{B_{\mathrm{H}}}{B_{\mathrm{H}}^{\mathrm{max}}} \right)^{-\frac{2p+3}{5}}  F_ {\mathrm{obs}}^{\frac{p+4}{5}} \, D_{\mathrm{s}}^{\frac{2(p+4)}{5}}   \left(\frac{M}{10^9M_{\odot}} \right)^{-\frac{2p+3}{10}}    \textrm{erg} \textrm{s}^{-1} ,
\label{pjet}
\end{split}
\end{equation}  
where 
\begin{equation}
\begin{split}
K_5 =  \: & \frac{\pi}{2} m_{\mathrm{p}} c^3 K_4^{-\frac{p+4}{5}} (5)^{\frac{2p+13}{5}} (2.96 \times 10^{14})^{-\frac{2p+13}{5}} \\
& (0.56 \times 10^{4})^{-\frac{2p+3}{5}} (0.05)^{\frac{p+4}{5}} ,
\end{split}
\end{equation} 
where $h = [(p+5)(p+4)]/10$ for single blob emission (Eq. \ref{blob}) or $h = [(p+3)(p+4)]/10$ for continuous blob emission (Eq. \ref{cblobs}) and $ f = f_0  k_0  (l_0  f_{\mathrm{ep}})^{-1}$.  We use a normalization value for the Lorentz factor of the jet, say 5, although this factor can range from $\sim$ 2 to $\sim$ 100, as observational data suggest. We adopt $f_{\mathrm{ep}} \sim 10^{-2} $ (and then $f_0 \simeq 1$), which means that there is, in average, one hundred electrons/positrons for at least one proton in a jet that is powered by a very rapidly spinning BH ($a_* \geqslant 0.95$) and observed at a large angle ($\geqslant 10^{\circ}$). Flat spectrum cores are predicted for any angle to the line of sight \citep{bk79}. They are pointing close to the line of sight only if the cores dominate over the extended emission.

Mildly-relativistic shocks ($1\lesssim \gamma_{\mathrm{s}} \lesssim 30$) are believed to occur in the AGN jets. \citet{ga99} have shown that at relativistic shocks, the particles typically perform a fraction $\sim \gamma_{\mathrm{s}}^{-1}$ of the Larmor orbit upstream before recrossing the shock, and the particle energy gain increases by a factor of $\gamma_{\mathrm{sh}}^{2}$ in the first shock-crossing cycle. In the subsequent shock-crossing cycles, the energy gain is of the order of 2. The predicted spectral index is $\simeq 2$. A series of Monte Carlo simulations (which are meant to find a way of constructing the trajectories of particles whose distribution obey the desired transport equation), performed under a wide range of background conditions at the shock front, indicate a value $p \simeq 2.2 - 2.3$ for the slope \cite[e.g.,][]{bo98,acht,kirk00,kw05}. It follows that in the regime of arbitrarily high Lorentz factor shocks, the acceleration process generates particle spectra which are quasi-independent of the considered background conditions, leading to a quasi-universal slope of $\sim 2.2$.  This picture can slightly be changed when one considers more realistic conditions in the vicinity of the shock. For example, \citet{jn06} studied possible models for perturbed magnetic field upstream of the shock and found that for superluminal, mildly-relativistic shocks, a flattening of the spectrum occurs, $p \simeq 1.5 $, with a cut off at lower energies than as expected for UHECRs. A flattening of the spectrum, $p \sim 2.1 - 1.5$, is also observed in simulation results obtained by \citet{meli} for the case of superluminal, ultra-relativistic shocks ($100 < \gamma_{\mathrm{sh}} < 1000$), with a turbulent magnetic field and various shock obliquity; however, for subluminal, mildly-relativistic shocks, the spectral slope has values between $2.0 < p < 2.3$.

For the following calculations, we adopt $p = 2$, which is the upper value of the spectral index for flat-spectrum core sources. The jet power (Eq. \ref{pjet}) then becomes:
\begin{equation}
\begin{split}
P_{\mathrm{j}} =  & 8.78 \times10^{36} \beta_{\mathrm{j}} \mathcal{D}_{\mathrm{j}}^{-3}
\left( \frac{\gamma_ {\mathrm{j}}}{5}\right) ^{17/5}\left( \frac{\tan\theta}{0.05}\right)^{6/5} \left( \frac{r_0}{2r_{\mathrm{g}}}\right) ^{-17/5}  \\ 
& \left(\frac{B_{\mathrm{H}}}{B_{\mathrm{H}}^{\mathrm{max}}} \right)^{-7/5} \left(\frac{F_ {\mathrm{obs}}}{\mathrm{mJy}} \right)^{6/5}\left( \frac{D_{\mathrm{s}}}{\mathrm{Mpc}}\right)^{12/5} \left(\frac{M}{10^9M_{\odot}} \right)^{-7/10},
\end{split}
\label{jet2}
\end{equation} 
where a continuous jet was considered $(h = 3)$. For $p\lesssim 2$, the results for the jet power are slightly reduced. 

\citet{bk79} showed that the flat-spectrum radio synchrotron emission of a compact jet core is produced by superposition of self-absorbed synchrotron spectra at different positions of the jet. In their model, the observed radio flux density depends on the jet power and the distance to the jet source $ D_{\mathrm{s}}$ as
\begin{equation}
F_ {\mathrm{obs}} \sim P_{\mathrm{j}}^{17/12}\, D_{\mathrm{s}}^{-2}.
\label{bk}
\end{equation} 

\citet{heinz-sunyaev} obtained a generalization of Eq. (\ref{bk}), for any scale-invariant jet model producing a power-law synchrotron spectrum with an index $\alpha$ in the form of $F_ {\mathrm{obs}} \sim P_{\mathrm{j}}^{(17+8\alpha)/12} M^{-\alpha}$, where $M$ is the BH mass. Because of the large mass difference between AGN and microquasars, this non-linearity function of the radio flux density with the BH mass indicates that the AGN jets are more radio-loud that the microquasar jets. 

\citet{fb95} found that for radio-loud\footnote{The radio-loudness parameter is defined as the ratio of radio luminosity emitted by the jet to the UV/X-ray luminosity emitted by the accretion disk.} AGN, the observed radio flux density depends non-linearly on the BH mass:
\begin{equation}
 F_ {\mathrm{obs}} \sim M^{1.42} \mathcal{D}_{\mathrm{j}}^{2.2}\, \gamma_ {\mathrm{j}}^{-1.8}.
\label{acc}
\end{equation} 
This result is obtained for the case of an accretion-dominated jet, $P_{\mathrm{j}} \sim L_{\mathrm{disk}}$, where $L_{\mathrm{disk}}$ is the luminosity of the disk \citep[see also][]{fb95b}.

Using Eq. (\ref{jet2}) for the observed radio flux density of a conical jet, we obtain:
\begin{equation}
F_ {\mathrm{obs}} \sim  P_{\mathrm{j}}^{5/6}\, D_{\mathrm{s}}^{-2}\,M^{7/6}\, \mathcal{D}_{\mathrm{j}}^{5/2}\, \gamma_ {\mathrm{j}}^{-17/6}(\tan\theta)^{-1},
\label{fpdm}
\end{equation} 
where the magnetic field along the jet varies as $B \sim \gamma_{\mathrm{j}}^{-1} z^{-1}$, the electron number density in the jet scales as $\sim \gamma_{\mathrm{j}} z^{-2}$, $B_{\mathrm{H}} \simeq B_{\mathrm{H}}^{\mathrm{max}}$, and $r_0 = 2\,r_{\rm g}$. Since the observed radio flux density in Eq. (\ref{fpdm}) is not dependent on the distance along the jet, the expression can be applied to microquasars as well.

\section{Luminosity and flux of the UHECR}
\label{lumUHECR}

In this section, we seek for the UHECR luminosity flux ($F_{\mathrm{CR}}$) specified as a function of the observed radio flux density. First, we consider the UHECR luminosity defined as
\begin{equation}
L_{\mathrm{CR}} =  \epsilon_{\mathrm{CR}} P_{\mathrm{j}} = \epsilon_{\mathrm{CR}} \gamma_ {\mathrm{j}}{\dot{M}}_{\mathrm{j}}c^2,
\label{lum}
\end{equation} 
where it is assumed that the UHECR luminosity is a fraction ($\epsilon_{\mathrm{CR}}$) of the jet power, with $P_{\mathrm{j}} = L_{\mathrm{kin}}+L_{\mathrm{magn}}+L_{\mathrm{CR}}$. If we were to adopt the point of view that the jet power is shared equally in a comoving frame between the baryonic matter, magnetic field, and cosmic rays extending to the highest energy, $\varepsilon_{\mathrm{CR}}\simeq 1/3$. In the jet-disk model of \citet{fb95}, the energy equipartition in the comoving frame appears to be a good approximation. It would also suggest that AGN driven by the BH spin-down power, and suffering from a low mass accretion rate, may attain a higher Lorentz factor, consistent with some observations.

Using the expression for the jet power (Eq. \ref{jet2}), the cosmic ray luminosity (Eq. \ref{lum}) becomes:
\begin{equation}
\begin{split}
L_{\mathrm{CR}} =  & 2.92\times10^{36}   \beta_{\mathrm{j}} \mathcal{D}_{\mathrm{j}}^{-3}
\left( \frac{\gamma_ {\mathrm{j}}}{5}\right) ^{17/5}\left( \frac{\tan\theta}{0.05}\right)^{6/5} \left( \frac{r_0}{2r_{\mathrm{g}}}\right) ^{-17/5}  \\ 
& \left(\frac{B_{\mathrm{H}}}{B_{\mathrm{H}}^{\mathrm{max}}} \right)^{-7/5} \left(\frac{F_ {\mathrm{obs}}}{\mathrm{mJy}} \right)^{6/5}\left( \frac{D_{\mathrm{s}}}{\mathrm{Mpc}}\right)^{12/5} \left(\frac{M}{10^9M_{\odot}} \right)^{-7/10}.
\end{split}
\label{lcrp2}
\end{equation} 

Given the UHECR luminosity, we can easily obtain the UHECR flux:
\begin{equation}
F_{\mathrm{CR}}=\frac{L_{\mathrm{CR}}}{4\pi D_{\mathrm{s}}^2},
\end{equation} 
where we do not include the cosmological distance as we refer to nearby radio, flat-spectrum core sources with a redshift up to $z \sim 0.018$.

\section{Maximum particle energy of the UHECR}
\label{emax}

Now, we look for the maximum energy of the UHECR in the case of the spatial (geometrical) limit \citep{fb95}; i.e., the jet particle orbits must fit into the Larmor radius. Conform to \citet{ga99}, the maximum particle energy in the downstream rest frame can be written as
\begin{equation}
E_{\mathrm{max}}^{\mathrm{sp}} =\gamma_{\mathrm{s}} e  Z B_0 r ,
\end{equation} 
where $\gamma_{\mathrm{s}}$ is the Lorentz factor of the shock and $Z$ is the particle mass number. Using the expression for the magnetic field along the jet (Eqs. \ref{magnfield2} and \ref{field}) and the fact that $\tan\theta=r_0/z_0$, the maximum energy of the UHECR particles (in the observer frame) becomes:
\begin{equation}
E_{\mathrm{max}}^{\mathrm{sp}} = e Z B_{\mathrm{H}}^{\mathrm{max}} r_0 \left( \frac{\gamma_{\mathrm{s}}}{\gamma_{\mathrm{j}}}\right) \left(\frac{B_{\mathrm{H}}}{B_{\mathrm{H}}^{\mathrm{max}}} \right).
\end{equation} 
For protons,
\begin{eqnarray}
E_{\mathrm{max}}^{\mathrm{sp}} = 5\times10^{20}   \left(\frac{B_{\mathrm{H}}}{B_{\mathrm{H}}^{\mathrm{max}}} \right) \left(\frac{r_0}{2r_\mathrm{g}} \right) \left( \frac{M}{10^9 M_{\odot}}\right)^{1/2} \; (\textrm{eV}),
\label{spatial}
\end{eqnarray} 
where $\gamma_{\mathrm{s}} \simeq \gamma_ {\mathrm{j}}$ was used.

Next, we look for the maximum energy of the UHECR in the case of the synchrotron loss limit \citep{bs}. Setting synchrotron losses equal to diffusive shock acceleration gains, \citet{bs} showed that a ubiquitous cutoff in the non-thermal emission spectra of AGN can be explained. This requires that the protons to be accelerated near $10^{21}$ eV. The frequency cutoff ($\nu_*$) might be produced at about $(3 - 6) \times10^{3} \,r_{\mathrm{g}}$. Rewriting the expression for the maximal proton energy derived by \citet{bs},
\begin{equation}
E_{\mathrm{max}}^{\mathrm{loss}} \simeq 1.4 \times 10^{20} \left( \frac{\nu_*}{3\times 10^{14}\mathrm{Hz}}\right)^{1/2} B^{-1/2}\; (\textrm{eV}) ,
\end{equation} 
and using the expression for the magnetic field along the jet (Eqs. \ref{magnfield2} and \ref{field}), the maximal proton energy in the loss limit reads:
\begin{equation}
\begin{split}
E_{\mathrm{max}}^{\mathrm{loss}}   \simeq  \: & 4.2\times 10^{18} \left( \frac{\nu_*}{3\times 10^{14}\mathrm{Hz}}\right)^{1/2}\left( \frac{\gamma_ {\mathrm{j}}}{5}\right) ^{1/2} \\ 
&\left(\frac{B_{\mathrm{H}}}{B_{\mathrm{H}}^{\mathrm{max}}} \right)^{-1/2}\left( \frac{M}{10^9 M_{\odot}}\right)^{1/2}\left( \frac{z}{z_0}\right)^{1/2}  \; (\textrm{eV}).
\end{split}
\label{loss}
\end{equation}

\section{Application to M87 and Cen A}
\label{m87cenA}

In this section, we investigate the UHECR luminosity flux for two possible sources of UHECR, M87 and Cen A, whose jet parameters can be inferred from observational data. 

Taking $r = 2\,r_{\mathrm{g}}$ and $B_{\mathrm{H}} \simeq B_{\mathrm{H}}^{\mathrm{max}}$, the equations for the maximum particle energy in the spatial (Eq. \ref{spatial}) and loss (Eq. \ref{loss}) limits, as well as for the UHECR luminosity (Eq. \ref{lcrp2}), become:
\begin{equation}
\begin{split}
E_{\mathrm{max}}^{\mathrm{sp}}   =  \: & 5\times10^{20}   \left( \frac{M}{10^9 M_{\odot}}\right)^{1/2} \; (\textrm{eV}) ,\\
E_{\mathrm{max}}^{\mathrm{loss}}  \simeq  \: & 4.2\times 10^{18} \left( \frac{\nu_*}{3\times 10^{14}\mathrm{Hz}}\right)^{1/2}\left( \frac{\gamma_ {\mathrm{j}}}{5}\right) ^{1/2} \\
& \left( \frac{M}{10^9 M_{\odot}}\right)^{1/2}  \left( \frac{z}{z_0}\right)^{1/2} \; (\textrm{eV}) , \\
L_{\mathrm{CR}}  =  \: & 3.65\times10^{38} \beta_{\mathrm{j}}  \left(1-\beta_{\mathrm{j}}  \cos\varphi \right)^{3}
\left( \frac{\gamma_ {\mathrm{j}}}{5}\right) ^{32/5}\left( \frac{\tan\theta}{0.05}\right)^{6/5}\\ 
&\left(\frac{F_ {\mathrm{obs}}}{\mathrm{mJy}} \right)^{6/5}\left( \frac{D_{\mathrm{s}}}{\mathrm{Mpc}}\right)^{12/5} \left(\frac{M}{10^9M_{\odot}} \right)^{-7/10} \; (\textrm{erg}\,  \textrm{s}^{-1}).
\end{split}
\label{toate}
\end{equation}
This set of equations is used for the following estimations on UHECR contribution from the BH spin-down power. For the expression of $E_{\mathrm{max}}^{\mathrm{loss}}$ in Eq. (\ref{toate}), we use $(z/z_0) \sim 10^3 $. Our choice is based on the results obtained by \citet{pbb09}, which show that a first large steady shock can be produced at about $z \sim 3\times10^3 \, r_{\mathrm{g}}$ [following the work by \citet{markoff01}]. This is confirmed by observations of a blazar inner jet as revealed by a radio-to-$\gamma$-ray outburst \citep{marscher}. The same conclusion was reached earlier by \citet{bs} using the observed cutoff in the radio emission of AGN.

Table \ref{Tb:M87CenA} contains our estimations for the maximum particle energy, luminosity, and flux of the UHECRs in the case of M87 and Cen A, whose jet parameters can be obtained from observational data. The observed radio flux density of the core corresponds to a frequency of 5 GHz. For comparison, we use the energies along the jet estimated by \citet{why-antonucci}, which are $ \sim 10^{45}$ erg s$^{-1}$ and $ \sim 10^{43}$ erg s$^{-1}$ for M87 and Cen A, respectively. The estimation of the luminosity and flux of the UHECR corresponds, however, to the upper limit of the slope of the particle density distribution of $p = 2$. For a steeper slope of the particle density distribution, the luminosity and flux of the UHECR increase, whereas for a flatter slope they decrease.
For instance, if we take $p = 2.4$ (so that $\alpha = 0.7$, the very low limit for a steep synchrotron emission spectrum), the luminosity and flux of the UHECR increase to values which are $\sim 2.71$ times those when $p = 2$ for M87 and $\sim 3.38$ for Cen A. 
Although both sources have a low jet power, they are powerful enough to provide the environment for particles to be accelerated to ultra-high energies of $\sim$ 1 ZeV. For both sources, the jet power is supplied by the BH spin-down power, as their mass accretion rate relative to the Eddington accretion rate is less than $\sim 10^{-2}$.

\section{Predictions for nearby galaxies as ultra-high-energy cosmic ray sources}
\label{sources}

We apply the model proposed here to a complete sample of steep spectrum radio sources \citep{biermann08,laur}, at redshift $z < 0.025$ (about 100 Mpc), with a total radio flux density larger than 0.5 Jy. The numbers for the estimated flux and maximal energy exclude the GZK effect, but includes the distance effect. The selection criteria used by the authors are presented in more detail in their papers. Table \ref{tab:sources} lists the predictions for the UHECR particle maximum energy, luminosity, and flux. We emphasize that there could be a common scaling limit, such as a condition that the Larmor radius has to fit three times or five times into the jet. The scaling limit is not critical to our predictions as long as we refer the quantities to, say, those of M87; therefore, the jet parameters are assumed to be the same as for M87, all scaled by BH mass, radio power, and distance to the source. This can be seen by comparing the estimated values of the UHECR maximum particle energy and flux for Cen A in Table \ref{Tb:M87CenA} (with the jet parameters inferred from observational data) and Table \ref{tab:sources} (with the jet parameters of M87); the differences are within one order of magnitude. Using a scaling relative to M87, one implicitly allows for the possibility that all sources produce higher nuclei ($Z > 1$). Of course, this again assumes that all sources are the same in this respect.  We argue that although the sources are LLAGN, they can be sites of accelerating particles to ultra-high energies $\sim$ 1 ZeV  with an UHECR luminosity $< 10^{43}$ erg s$^{-1}$. 

\begin{figure}\centering
\epsfig{file=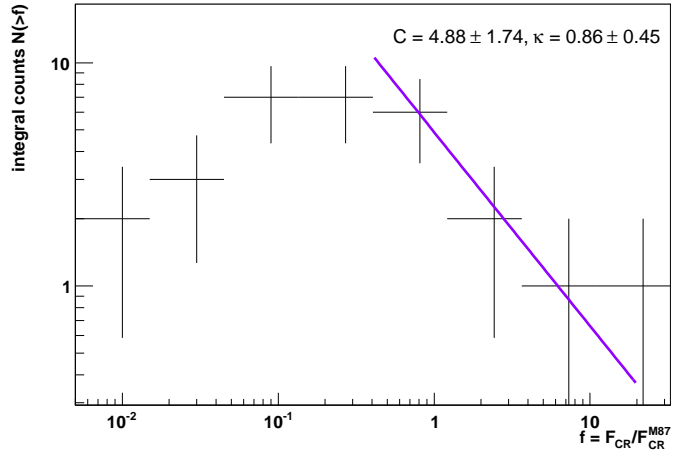,height=6cm}
\caption{Integral source counts of the same sample in Table \ref{tab:sources}. The model parameters for the best fit of a power-law model ($N = C\, f^{\kappa}$) with $N(>0.405)$ are: $C = 4.88 \pm 1.74$ and $\kappa = 0.86 \pm 0.45$.}
\label{histoFit}
\end{figure}

In Fig. \ref{histoFit}, we show the plot of integral source counts ($N$) versus relative UHECR flux (last column in Table \ref{tab:sources}), where the data points are split in 8 bins of variable width with the bin upper edge three times larger than the bin lower edge. Although the data points are too sparse to discriminate models of the flux distribution, we estimate the most likely model parameters that could generate these data points. We fit a power-law model ($N = C\, f^{\kappa}$) to the mean values of the data in each bin for $N(>0.405)$ using the log-likelihood method.\footnote{To plot numerical results and to analyze data, we use ROOT, which is an open source, data analysis program developed at CERN.} The errors for the data points are estimated by $\sqrt{N}$. For the best fit that we obtained, the model parameters are: $C = 4.88 \pm 1.74$ and $\kappa = 0.86 \pm 0.45$. (Since the source counts for each bin is too low, especially in the tail of the assumed power-law distribution, we cannot perform a $\chi^2$-test.) When $\kappa < 1$, the contribution to the UHECR flux from a few strong radio sources dominates over that from many weak radio sources. On the other hand, within the errors we can also consider a power-law model with $\kappa > 1$, which implies that the contribution to the UHECR flux from many weak radio sources is dominant. To distinguish one possibility from the other one, more information on radio data and BH mass from other nearby AGN, in additional to those listed in Table \ref{tab:sources}, are required; so that, a larger complete sample of nearby AGN may lead to an improved statistical model to answer the question of whether the flux of the UHECRs is produced by many weak radio sources or a few strong radio sources.

\section{Summary and conclusions} 
\label{sec:summary}

In this work we developed a new model for UHECR contribution from BH spin-down power. We relate the observed radio flux density to the luminosity and flux of the UHECRs, and calculate the maximum particle energy in both spatial and loss limits. 

We can attribute the production of UHECRs in LLAGN based on the fact that in LLAGN the jet power can be magnetically provided by the BH spin-down power, which is dependent on the square of the strength of the BH magnetic field ($P_{\rm j} \sim B_{\rm H}^2$). Thus, the particle acceleration regions can be sustained by the magnetic energy extraction from spinning BHs, where the strength of the magnetic field at the sites of particle acceleration scales with the maximum value of the BH magnetic field, which is $\sim 10^4$ gauss for a BH of $10^9 M_{\odot}$, and where the maximum particle energy is given by Eqs. \ref{spatial} and \ref{loss}.

The model proposed here is relevant for many nearby radio galaxies. Since their accretion rate is very low, the BH spin-down power, which is a better option than the accretion power, can sustain the required conditions for particle acceleration along the jets. Our predictions can be used in various propagation codes for UHECRs, which typically use the maximal energy and flux of the cosmic rays scaled relative to a canonical radio galaxy (e.g., columns 8 and 11 in Table 2), as well as the slope of the CR spectrum. The only difficulty is that the predicted quantities scale a slightly differently if some of the UHECRs are heavy nuclei at the origin. That is because for heavy nuclei, the energy scales up with the atomic number. For a middling FIR/Radio ratio ('FIR' is the flux density at 60 $\mu$ and 'Radio' is the flux density at 5 GHz), as Cen A has, a starburst may have provided all the necessary pre-accelerated nuclei. Nonetheless, this is not the case for M87. 

More information on radio data and BH mass from other nearby AGN are required to develop an improved statistical model which may answer the question of whether the flux of the UHECRs is produced by many weak radio galaxies or a few strong radio galaxies.

\vspace{1cm}
\begin{small}
The author would like to thank Peter L. Biermann. The author would also like to thank Lauren\c{t}iu I. Caramete for providing her with a complete sample of AGN. This research was supported through a stipend from the International Max Planck Research School (IMPRS) for Astronomy and Astrophysics at the Universities of Bonn and Cologne. The author appreciates the support from MPIfR during the last phase of this work.
\end{small}

\newpage

\begin{landscape}
\begin{table*}
\centering
\caption{The jet parameters for M87 and Cen A and the corresponding estimation of the maximum particle energy (spatial and loss limits), luminosity, and flux of the UHECRs, were we assume that the particles are protons.}\label{Tb:M87CenA}
\begin{threeparttable}
\begin{tabular}{c c c c c c c c c c c c}
\hline \hline
Source & $\gamma_{\mathrm{j}}$ & $\varphi$ & $\theta$ & $D_{\mathrm{s}}$  & $M$  & $F_ {\mathrm{core}}^{5\mathrm{GHz}}$ & $E_{\mathrm{max}}^{\mathrm{sp}}$ & $E_{\mathrm{max}}^{\mathrm{loss}}$ & $L_{\mathrm{CR}}$ & $F_{\mathrm{CR}}$ & $L_{\mathrm{jet}}$ \\
& &($\circ$)& ($\circ$) & (Mpc) & $(\times10^9M_{\odot})$ &  (mJy) & ($10^{20}$ eV) & ($10^{20}$ eV) & (erg s$^{-1}$)& (erg s$^{-1}$cm$^{-2}$)& (erg s$^{-1}$) \\[0.5ex]
\hline
M87 & 6 & 10 & 19 & 16 & 3 & 2875.1  & 8.61 & 2.50 &  $2.03\times10^{43}$& 7.03$\times10^{-10}$   & $ \sim 10^{45}$ \\
Cen A & 2 & 65 & 5 & 3.5 & 0.055 &  6984 &  1.16 & 2.17 & $ 3.16\times10^{42}$ & 2.28$\times10^{-9}$  &  $ \sim 10^{43}$ \\
\hline
\end{tabular}
\scriptsize
\hfill\parbox[t]{20.2cm}{NOTE: Col. (1) Source name; Col. (2) Jet Lorentz factor; Col. (3)  Angle to the line of sight; Col. (4) Jet semi-opening angle; Col. (5) Distance to the source; Col. (6) BH mass; Col. (7) Core flux density at 5 GHz; Col. (8) maximum particle energy (spatial limit); Col (9) maximum particle energy (loss limit); Col. (10) UHECR luminosity for a power-law index of particle distribution $p =2$; Col. (11) UHECR flux; Col. (12) Estimated energy flow along the jet $L_{\mathrm{jet}}$. REFERENCES: Jet Lorentz factor: \citet{biretta99} and \citet{meisenheimer}; Angle to the line of sight: \citet{biretta99} and \citet{tingay}; Distance to the source for M87 \citet{macri} and for Cen A we assume a distance $D_{\mathrm{s}} = 3.5$ Mpc to be consistent with the BH mass determination; BH mass: \citet{macchetto} and \citet{cappe}; Core flux density at 5 GHz: \citet{nagar} and \citet{slee}; Estimated energy flow along the jet $L_{\mathrm{jet}}$: \citet{why-antonucci}. The value of the quantities inferred from observational data are the median ones, which we use in our estimations.}
\end{threeparttable}
\end{table*}
\end{landscape}

\begin{landscape}
\begin{table}
\centering
\caption{UHECR predictions for a complete sample of 29 steep spectrum sources \citep{biermann08,laur}.}\label{tab:sources}
\begin{threeparttable}
\begin{tabular}{l  c c c c c c c c c c c c c c}
\hline
\hline
Source & D & M & $F_{\mathrm{core}}^{5\mathrm{GHz}}$ & $ E_{\mathrm{max}}^{\mathrm{sp}} $ & $ E_{\mathrm{max}}^{\mathrm{sp}} $/ & $ E_{\mathrm{max}}^{\mathrm{loss}} $ & $ E_{\mathrm{max}}^{\mathrm{loss}} $/ &$ L_{\mathrm{CR}}$ & $ F_{\mathrm{CR}}$ &$ F_{\mathrm{CR}}$/ \\

& (Mpc)& ($\times10^9 M_\odot $)&(mJy)& ($10^{20}$ eV)   &    $E_{\mathrm{max}}^{\mathrm{sp,M87}}$ & ($10^{20}$ eV)   &    $E_{\mathrm{max}}^{\mathrm{loss,M87}}$ & (erg s$^{-1}$) & (erg s$^{-1}$ cm$^{-2}$) &  $F_{\mathrm{CR}}^{\mathrm{M87}}$   \\

\begin{small}(1)\end{small}&\begin{small}(2)\end{small}&\begin{small}(3)\end{small}&\begin{small}(4)\end{small}&\begin{small}(5)\end{small}&\begin{small}(6)\end{small}&\begin{small}(7)\end{small}&\begin{small}(8)\end{small}&\begin{small}(9)\end{small}&\begin{small}(10)\end{small}&\begin{small}(11)\end{small}\\[0.5ex]
\hline
ARP 308&69.7&0.1&67.8&1.57&0.18&0.45&0.18&8.39$\times10^{43}$&1.52$\times10^{-10}$&0.21\\
CGCG 114-025&67.4&0.19&443.39&2.16&0.25&0.63&0.25&4.70$\times10^{44}$&9.15$\times10^{-10}$&1.30\\
ESO 137-G006&76.2&0.92&631.32&4.76&0.55&1.38&0.55&3.19$\times10^{44}$&4.87$\times10^{-10}$&0.69\\
IC 4296&54.9&1&214&4.97&0.57&1.44&0.57&3.75$\times10^{43}$&1.10$\times10^{-10}$&0.15\\
IC 5063&44.9&0.2&29&2.22&0.25&0.64&0.25&6.49$\times10^{42}$&2.84$\times10^{-11}$&0.04\\
NGC 0193&55.5&0.2&40&2.22&0.25&0.64&0.25&1.58$\times10^{43}$&4.55$\times10^{-11}$&0.06\\
NGC 0383&65.8&0.55&92&3.68&0.42&1.07&0.42&3.19$\times10^{43}$&6.52$\times10^{-11}$&0.09\\
NGC 1128&92.2&0.2&39&2.22&0.25&0.64&0.25&5.20$\times10^{43}$&5.41$\times10^{-11}$&0.07\\
NGC 1167&65.2&0.46&243&3.37&0.39&0.98&0.39&1.13$\times10^{44}$&2.36$\times10^{-10}$&0.33\\
NGC 1316&22.6&0.92&26&4.76&0.55&1.38&0.55&3.76$\times10^{41}$&6.51$\times10^{-12}$&0.01\\
NGC 1399&15.9&0.3&10&2.72&0.31&0.79&0.31&1.12$\times10^{41}$&3.94$\times10^{-12}$&0.005\\
NGC 2663&32.5&0.61&160&3.88&0.45&1.13&0.45&1.06$\times10^{43}$&8.89$\times10^{-11}$&0.12\\
NGC 3801&50&0.22&635&2.33&0.27&0.67&0.27&3.19$\times10^{44}$&1.12$\times10^{-9}$&1.60\\
NGC 3862&93.7&0.44&1674&3.29&0.38&0.96&0.38&2.83$\times10^{45}$&2.85$\times10^{-9}$&4.06\\
NGC 4261&16.5&0.52&390&3.58&0.41&1.04&0.41&6.80$\times10^{42}$&2.20$\times10^{-10}$&0.31\\
NGC 4374&16&1&168.7&4.97&0.57&1.44&0.57&1.46$\times10^{42}$&5.05$\times10^{-11}$&0.07\\
NGC 4486&16&3&2875.1&8.61&1&2.50&1&2.03$\times10^{43}$&7.03$\times10^{-10}$&1\\
NGC 4651&18.3&0.04&15&0.99&0.11&0.28&0.11&1.05$\times10^{42}$&2.78$\times10^{-11}$&0.03\\
NGC 4696&44.4&0.3&55&2.72&0.31&0.79&0.31&1.02$\times10^{43}$&4.59$\times10^{-11}$&0.06\\
NGC 5090&50.4&0.74&268&4.27&0.49&1.24&0.49&4.94$\times10^{43}$&1.72$\times10^{-10}$&0.24\\
NGC 5128&3.5&0.055&6984&1.16&0.13&0.33&0.13&2.53$\times10^{43}$&1.82$\times10^{-8}$&25.95\\
NGC 5532&104.8&1.08&77&5.16&0.6&1.50&0.6&4.92$\times10^{43}$&3.96$\times10^{-11}$&0.05\\
NGC 5793&50.8&0.14&95.38&1.86&0.21&0.54&0.21&4.67$\times10^{43}$&1.60$\times10^{-10}$&0.22\\
NGC 7075&72.7&0.25&20&2.48&0.28&0.72&0.28&1.13$\times10^{43}$&1.89$\times10^{-11}$&0.02\\
UGC 01841&84.4&0.1&182&1.57&0.18&0.45&0.18&4.34$\times10^{44}$&5.39$\times10^{-10}$&0.76\\
UGC 02783&82.6&0.42&541&3.22&0.37&0.93&0.37&5.58$\times10^{44}$&7.23$\times10^{-10}$&1.02\\
UGC 11294&63.6&0.29&314&2.67&0.31&0.78&0.31&2.01$\times10^{44}$&4.39$\times10^{-10}$&0.62\\
VV 201&66.2&0.1&88&1.57&0.18&0.45&0.18&1.01$\times10^{44}$&2.04$\times10^{-10}$&0.29\\
WEIN 45&84.6&0.27&321.6&2.58&0.3&0.75&0.3&4.31$\times10^{44}$&5.33$\times10^{-10}$&0.75\\
\hline
\end{tabular}
\end{threeparttable}
\end{table}
\end{landscape}
\scriptsize
NOTE: Col. (1) Source name; Col. (2) Distance to the source; Col. (3) BH mass; Col. (4) Core flux density at 5 GHz; Col (5) Maximum particle energy (spatial limit); Col. (6) Maximum particle energy relative to that of M87; Col. (7) UHECR luminosity; (8) UHECR flux; (8) UHECR flux relative to that of M87. The value of the quantities inferred from observational data are the median ones, which we use in our estimations.

\end{document}